# Ultralarge polarization in ferroelectric hafnia-based thin films

Han Wu[1], Kun Lin[1], Qinghua Zhang[2], Qian Yu[3], Xiaoqian Fu[3,6], Qiang Li[1], Meera Cheviri[4], Oswaldo Diéguez[4], Shuai Xu[2], Lin Gu[2], Yili Cao[1], Jiaou Wang[5], Zhen Wang[5], Yu Chen[5], Huanhua Wang[5], Jinxia Deng[1], Jun Miao[1], Xianran Xing[1,*]

1 Beijing Advanced Innovation Center for Materials Genome Engineering, Institute of Solid State Chemistry, University of Science and Technology Beijing, Beijing 100083, China.

2 Beijing National Laboratory for Condensed Matter Physics, Institute of Physics, Chinese Academy of Sciences, Beijing 100190, China.

3 Center of Electron Microscopy and State Key Laboratory of Silicon Materials, School of Materials Science and Engineering, Zhejiang University, Hangzhou, 310027, China.

4 Department of Materials Science and Engineering, The Iby and Aladar Fleischman Faculty of Engineering, and The Raymond and Beverly Sackler Center for Computational Molecular and Materials Science, Tel Aviv University, Tel Aviv, Israel.

5 Institute of High Energy Physics, University of Chinese Academy of Sciences, Chinese Academy of Sciences, Beijing 100049, China.

6 Pico Electron Microscopy Center, School of Materials Science and Engineering, Hainan University, Haikou, Hainan 570228, China.

* Corresponding author. Email: xing@ustb.edu.cn (X.X.)

**Abstract**

Hafnia-based ferroelectrics have become a valuable class of electronic functional materials at the nanoscale, showing great potential for next-generation memory and logic devices. However, more robust ferroelectric properties and better understanding of the polarization mechanisms are currently needed both in technology and science. Herein, we report the properties of oxygen-deficient $Hf_{0.5}Zr_{0.5}O_2$ films with ultralarge remanent polarization ($P_r$) of 387 μC cm$^{-2}$ at room temperature (1 kHz). Structure characterizations identify a new ferroelectric monoclinic $P$c phase in these $Hf_{0.5}Zr_{0.5}O_2$ films. The in-situ STEM measurements evidence polar displacements of the oxygen atoms, which move up and down in the $P$c structure under applied DC bias fields, showing a huge displacement ($\triangle\delta= 1.6$ Å). DFT calculations optimized the $P$c structure and also predicted a large polarization. The coexistence of the ferroelectric monoclinic ($P$c) phases and orthorhombic ($P$ca2$_1$) is responsible for this superior ferroelectric properties. These findings are promising for hafnia-based ferroelectric applications in integrated ferroelectric devices, energy harvesting and actuators, etc.

**Introduction**

Hafnia-based ferroelectrics are of high technological and industrial significance, mainly due to their compatibility with complementary metal-oxide semiconductor (CMOS) processes and low-temperature synthesis characteristics (*1, 2*). Meanwhile, robust ferroelectric properties with nanoscale dimensions has been demonstrated, even at a lateral dimension of one nanometer, irrespective of depolarization fields (*3*). Hafnia possesses a structural polymorphism, the monoclinic phase ($P2_1$/c) at room temperature, tetragonal phase ($P4_2$/nmc) above 2100K, cubic phase ($F$m-3m) above 2800K, and orthorhombic phase ($P$ca2$_1$) and rhombohedral phase ($R$3m) in the doped epitaxial hafnia thin films (*4*). Currently, the orthorhombic phase ($P$ca2$_1$) is accounted for the primary contributor to the ferroelectricity. Many research groups have been motivated to enhance the spontaneous polarization ($P_S$) of hafnia-based thin films by different methods and synthesis techniques. Introducing oxygen vacancies via helium ion bombardment into $HfO_2$ thin films has been utilized to increase the molar volume and decrease the formation energy of $Pca2_1$ phase, resulting in an increase of $P_r$ from 5.5 µC cm$^{-2}$ to 7.5 µC cm$^{-2}$ (*5*). Particularly, an interface engineering strategy was employed to control the termination of the bottom (001) $La_{0.67}Sr_{0.33}MnO_3$ layers, raising the $P_r$ of (111)-oriented $Hf_{0.5}Zr_{0.5}O_2$ thin films from 10 µC cm$^{-2}$ to 20 µC cm$^{-2}$ (*6*). Doping with yttrium and maximizing grain sizes have also been employed to prepare hafnia-based films and a high $P_r$ (50 µC cm$^{-2}$) has been obtained (*7*). Intriguingly, a ferroelectric rhombohedral phase was found with a large $P_r$ of 34 µC cm$^{-2}$ (*8*). Up to now, however, the intrinsic origin of ferroelectricity in this material remains poorly understood, due to its unique polymorphism, complicated structures and dependence on defects (*4*). Compared with conventional perovskite ferroelectrics, the chemical diversity and tunability of the ferroelectric response still are challenges. Theoretic research has predicted that further enhancing the polarization of $Pca2_1$ phase entails excessively large epitaxial strain (~7%) (*9*), which is a barrier for films preparation. Moreover, the fully textured [001] (the direction of largest polarization of $Pca2_1$ phase) hafnia-based thin films commonly suffer from non-negligible leakage current, accompanied by a degradation of ferroelectric response (*10*).

In this work, we show the oxygen-deficient $Hf_{0.5}Zr_{0.5}O_2$ (HZO) thin films with boosted ferroelectric polarization, which were grown by pulsed laser deposition (PLD) on (001) $SrTiO_3$ (STO) substrate with the $La_{0.67}Sr_{0.33}MnO_3$ (LSMO) serving as conductive buffered layer, subjected to a long-range post-annealing process. The ultralarge $P_r$ of 387 µC cm$^{-2}$ at room temperature and 1 kHz have been achieved in the 10-nm-thickness HZO thin films. These results are of considerable importance to hafnia-based ferroelectrics.

**HZO thin films and structure characterization**

HZO thin films were grown on LSMO-buffered (001)-STO substrates using pulsed laser deposition (PLD), annealed at 750 °C for 240 minutes under ambient atmosphere (Supplementary materials: thin film preparation and fig.S1). Figure 1A shows the typical microstructure for a 10-nm-thickness annealed HZO thin film, imaged by high-angle annular dark-field scanning transmission electron microscope (HAADF-STEM). One can clearly observe the existence of laterally intertwined multiple domains with a size of ~10 nm (marked by dashed line). The surface of HZO, and the interfaces of HZO/LSMO and LSMO/STO are exceedingly flat, in consistent with Laue fringes and oscillating peaks observed in synchrotron x-ray diffraction (SXRD) and X-ray reflectivity (XRR), respectively. The XRR profile presents the regular thickness oscillations, displaying a high uniformity of layer thickness (Fig. 1B and fig. S2). The SXRD patterns exhibit a polymorphism character, the strongest diffraction peaks STO (001) and LSMO (001) assigned to STO substrates and epitaxially grown 10-nm-thickness LSMO layers, respectively, and the Bragg peaks around 30.3° assigned to the (111) reflections of $P$ca2$_1$ phase (Fig. 1C and fig. S3). However, it is worth noting that the emergent Bragg peak around 34.5° has been primary indexed as the (002) reflection of the $P$c structure, a new phase which has not been reported in hafnia-based thin films. The (002) reflection shifts to higher angles upon annealing in air, implying lattice shrinkage caused by the reduction of oxygen vacancy concentration (*11*). The presence of oxygen vacancies has well been confirmed by x-ray photoemission spectroscopy (XPS) (fig. S4). Moreover, the annealed thin films possess a larger crystal field splitting energy with respect to that of the unannealed counterparts (fig. S5), which is particular to improve the ferroelectric behaviors.

The detected coexistence of $P$ca2$_1$ and $P$c phases in the HZO thin films could be confirmed by second harmonic generation (SHG) measurements, which are performed to investigate the inversion symmetry breaking of thin films. For comparison, the SHG signals on LSMO thin films epitaxially grown on (001) STO with a thickness of 10 nm initially were assessed and the results of $I^{2\omega}_{p-out}$ and $I^{2\omega}_{s-out}$ were indicative of a tetragonal symmetry (4mm) (Fig. 1D). Furthermore, the HZO thin films exhibit more robust SHG signals under identical laser power (Fig. 1E and fig. S7), demonstrating highly asymmetric structures.

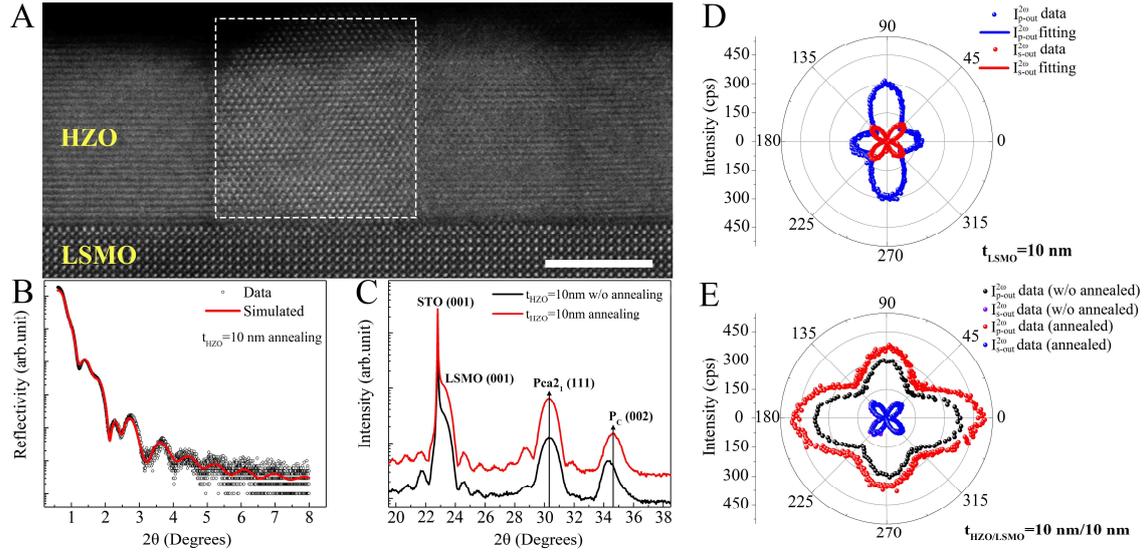

**Fig. 1. Morphology and structure.** (**A**) Representative cross-sectional HAADF-STEM image of 10-nm-thickness annealed HZO thin films. Scale bar, 5 nm. White dashed line indicates one of the domains. (**B**) XRR data and the corresponding fitting result of the HZO thin films. (**C**) Synchrotron XRD patterns of the annealed (red) and unannealed (black) samples, respectively. Curves are offset vertically for clarity. The vertical solid lines indicate the position of the diffraction peaks. (**D** and **E**) The polarization-angle-dependent SHG characterizations of the 10-nm-thickness LSMO thin film and HZO thin films, respectively.

**Ferroelectric properties of HZO thin films**

To test the ferroelectric switchable behavior of HZO thin films, we performed polarization versus voltage (P-V) loops through positive up negative down (PUND) measurements. Bistable switching and hysteresis loops at room temperature and a frequency of 1 kHz could be observed for the 10-nm-thickness annealed HZO thin films (Fig. 2A). The current versus voltage (I-V) curve distinctly demonstrates ferroelectric switching signals. The small gap and asymmetry of the imprint shift in P-V loops could be ascribed to the defects, like charged oxygen vacancies, aggregating under electric filed and leading to a built-in electric field (*12-14*).

Notably, an ultralarge spontaneous polarization $P_r$ of 387 μC cm$^{-2}$ has been achieved, which is an order of magnitudes greater than those of hafnia-based oxides reported previously (*15-19*). To our knowledge, the highest $P_r$ was 50 μC cm$^{-2}$ in yttrium-doped HfO$_2$ films so far (*7*). The $P_r$ is even much larger than those of conventional perovskite ferroelectric thin films, such as strained $BaTiO_3$ (*20*), $BiFeO_3$ (*21*) and $PbTiO_3$ (*22*). Figure 2B depicted the comparison of the largest $P_r$ of conventional ferroelectrics with this data. Besides, the coercive voltages ($V_C$) are merely ~2.2 V (coercive field $E_C \approx 2.3$ MV cm$^{-1}$) in the present case. Low coercive voltage signifies low switching energy of ferroelectric thin films. Achieving a low coercive voltage with a high $P_r$ are crucial in the pursuit of ultralow-power ferroelectric memory and logic devices (*23*).

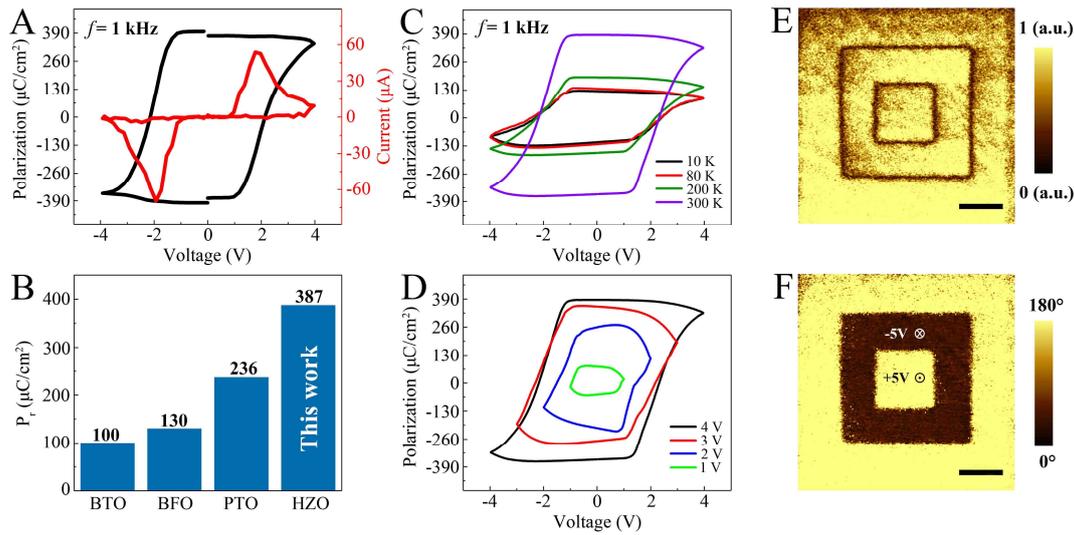

**Fig. 2. Ferroelectric characterization.** (**A**) The PUND measurements of a 10-nm-thickness annealed HZO thin film under external applied voltage with a frequency of 1 kHz at room temperature. The red line represents the current (I) versus voltage (V) curve after excluding the non-ferroelectric switching response; the black line represents the P–V hysteresis loops by integrating the corresponding ferroelectric switching current over time. (**B**) $P_r$ comparison of the present result with classical perovskite ferroelectric thin films. (**C**) P–V loops with temperatures, and (**D**) P–V loops with applied voltages at 1 kHz in the 10-nm-thickness annealed HZO thin films. (**E and F**) Amplitude (E) and phase (F) of the PFM images on a 10-nm-thickness annealed HZO thin film after poling with +5 V and –5 V. Scale bar, 1 μm.

Such superior intrinsic ferroelectric property has been further confirmed by temperature-dependent P-V loop measurements. Figure 2C reveals a distinct temperature dependence of the P-V loops for a 10-nm-thickness annealed HZO thin film from 300 K to 10 K. Remarkably, $P_r$ decreased from 390 μC cm$^{-2}$ at 300 K to 125 μC cm$^{-2}$ at 10 K, contrary to the conventional perovskite ferroelectrics with raised $P_r$ on cooling (*24-26*). Based on the pronounced change in $P_r$ with temperature, we propose that oxygen vacancies play a decisive role in enhancing polarization. Recent published research has reported that the inherent $P_r$ was estimated to be less than 9 µC cm$^{-2}$ in (111) HZO/ (001) LSMO system, but the measured value could reach 35 µC cm$^{-2}$ with the contribution of oxygen vacancy migration (*27*). Additionally, the $P_r$ in HZO thin films grown on LSMO/LaNiO$_3$/CeO$_2$/YSZ/Si has been shown to drop dramatically from 280 K to 33 K (*28*). In our case, the $P_r$ stays nearly constant below 80 K. This suggests that the internal mobile charged defects (e.g., charged oxygen vacancies) have already been 'frozen-in' at low temperatures (*29, 30*), leading to the negligible contributions of mobile charged oxygen vacancies to the $P_r$. Subsequently, the reproducibility of the ultralarge polarization $P_r$ (around 400 µC cm$^{-2}$) has been confirmed by P-V loops measurements with different batches of 10-nm-thickness annealed HZO thin films prepared in the same conditions (fig. S9). Polarization switching behaviors with increasing external voltage manifest a switching character of the escalating polarization correlated with the progressive switching of polar domains under increasing external voltage (Fig. 2D and fig. S10).

The prepared HZO thin film with excellent local ferroelectric response demonstrates the robust piezoresponses, which have been investigated by the piezoresponse force microscopy (PFM). The amplitude image exhibits an intensive piezoelectric response, with the zero amplitude regions being designated as domain walls due to the lack of polarization components (Fig. 2E). The corresponding phase image shows well-defined domain patterns with 180° phase contrast, representing the upward and downward remanent polarization states, confirming the robust ferroelectric property (Fig. 2F).

**STEM observations and a new ferroelectric monoclinic *P*c Phase**

SHG and STEM experiments show a laterally-intertwined domains distribution in the 10-nm-thickness HZO thin films (Fig. 1A and 1E). Further HAADF-STEM images reveal the fact of two phases coexistence, [111]-oriented *P*ca2$_1$-phase domains and [001]-oriented *P*c-phase domains (fig. S11). The former has helical domain structure (due to the four-fold rotational symmetry of (001) LSMO bottom layers). The latter, however, presents Hf/Zr arrangement that is highly similar to that of $P2_1/c$ phase ($a \approx 5.08$ Å, $b \approx 5.17$ Å, $c \approx 5.25$ Å).

More details are depicted in the typical HAADF-STEM images along [110], [100]

and [010] zone axes in Figure 3A, with magnified regions shown in Figure 3B. The ABF-STEM images can distinguish the oxygen atoms as shown in Figure 3C. It is intriguing to find that the oxygen columns demonstrate alternately dark and bright distribution along lateral directions. This differs remarkably from the $P2_1/c$ phase, where oxygen columns are expected to be uniformly distributed. Moreover, the present monoclinic $Pc$ phase adopt the diffraction spots (010) in FFT patterns (fig. S11C and 11F), but which disappeared in that of the $P2_1/c$ structure (fig. S12). These results, including the Bragg peaks at ~34.5° in Figure 1C, are indicative of a new monoclinic $Pc$ phase in the prepared HZO thin films.

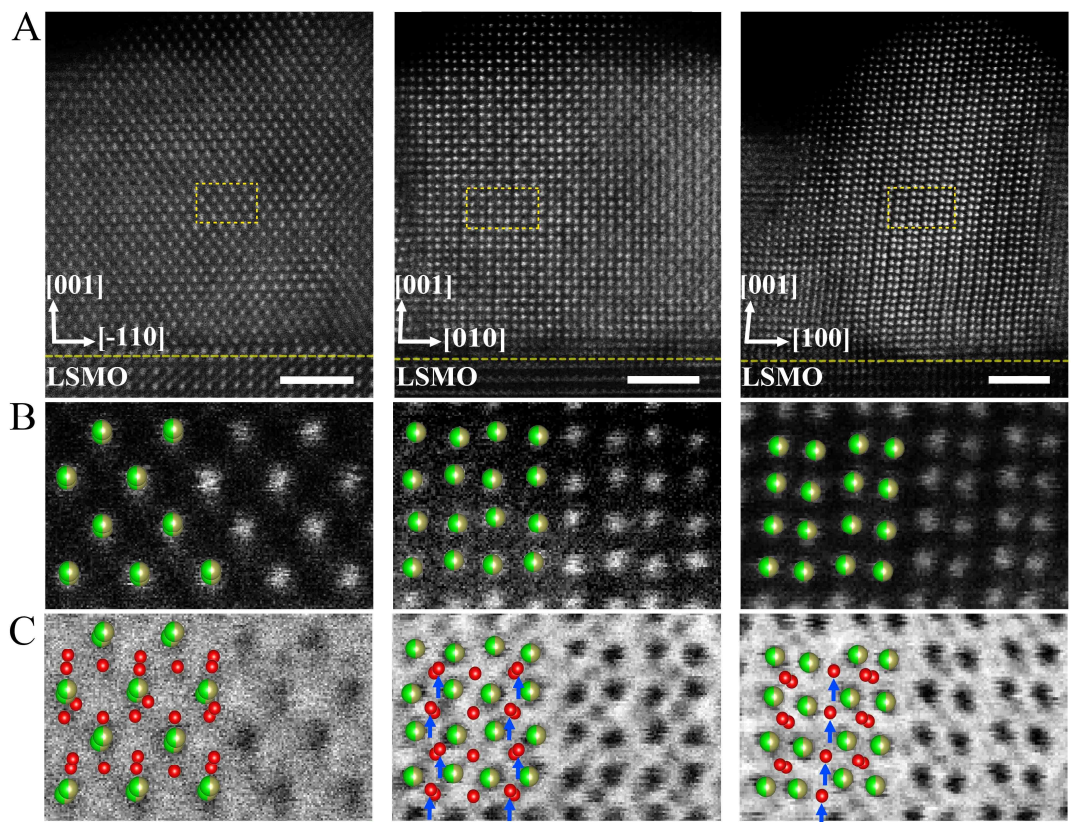

**Fig. 3. Atomic-scale structural analysis.** (**A**) Three representative HAADF-STEM images assigned for $Pc$-domains. The nanograins were aligned to the crystal zone axis for atomic resolution imaging. Scale bar, 2 nm. (**B**) Enlarged atomic structure images of the regions marked by yellow dashed lines in (A). The Hf/Zr atom models (green-yellow balls) were overlaid on each image for comparison. (**C**) Corresponding ABF-STEM images of (B). Oxygen atom models (red balls) were used for comparison. The blue arrows schematically denoted the position of off-center oxygen atoms.

We reconstructed the crystal structure of this new monoclinic phase based on the STEM images along three different directions ([110], [100] and [010] zone axes), and finally obtained a primitive structure with favorable consistency. In comparison to $P2_1/c$ phase, 1/4 oxygen atoms are lost in an ordered manner along the [001] direction in the primitive structure, leaving the local chemical composition to be $HfZrO_3$. This agrees with our XPS results with lower Zr 3d binding energy (fig. S4). The primitive structure was further optimized by first-principles calculations and reasonably identified as a monoclinic $Pc$ phase. The ABF-STEM images along both the [100] and [010] zone axes witness these off-centered oxygen atoms, denoted by blue arrows (Figure 3C).

The polarization switching of ferroelectric $Pc$ phase has been convinced by the in-situ electric bias STEM to observe the shifts of the off-center oxygen atoms. As shown in Figure 4C and 4D, the off-center oxygen atoms move down and up, upon being polarized by DC bias fields, which correspond to polarization down and up states, respectively. The polarization down and up states have been shown in Figure 4E and 4F. We are surprised to find the huge displacement of off-center oxygen atoms (1.6 Å) with up and down movements. The inherent $P_r$ of $Pc$ phase can be estimated around 110 μC cm$^{-2}$, slightly different from the measured value (125 μC cm$^{-2}$) of 10-nm-thickness annealed HZO films at 10K, owing to the contribution of (111) $Pca2_1$-phase components. These results provide direct evidences that the inherent polarization switching in the present HZO thin films mainly results from the shifts of off-center oxygen atoms in the $Pc$ phase and coordinated effect with $Pca2_1$ phase.

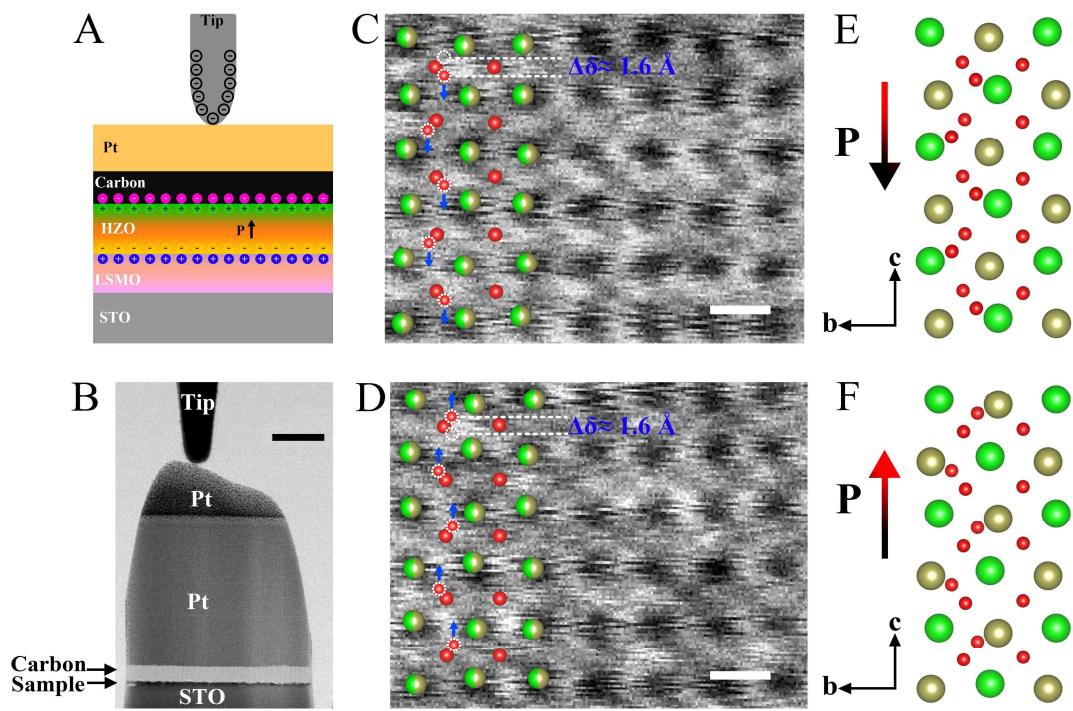

**Fig. 4. In-situ electric bias STEM tests.** (**A**) Schematic illustration for charge distribution after the FIB sample lamella poled by bias. The LSMO layers effectively supply sufficient free carriers to screen the polarization charge of HZO, facilitating the full polarization of HZO layers. (**B**) STEM image of the contact between a conductive tungsten probe and the FIB sample lamella. Scale bar, 200 nm. (**C and D**) ABF-STEM images of the uniformly polarized states of 10-nm-thickness HZO thin films in the (100) plane after applying in-situ forward and reverse electric bias. The blue arrows schematically denoted the moving direction of off-center oxygen atoms under electric bias. Scale bar, 0.25 nm. (**E and F**) Computed structures with forward and reverse directions of polarization, corresponding to C and D, respectively. Polarization vectors are represented by red-black arrows at left.

## Theoretical calculations

To shed light into the properties of our films, we performed density-functional theory calculations (see more details about our methodology in the Supplementary Information). Our starting point was the monoclinic structure of the $P2_1/c$ phase of $HfO_2$ and $ZrO_2$ (similar to Figure 5A), created using unit cells with 4 metal atoms and 8 oxygen atoms. From those cells, we built oxygen-deficient hafnia and zirconia by removing two oxygen atoms, which we then optimized by allowing the cell vectors and atomic positions to reach minimum energies (Fig. S13). These optimizations resulted in several phases, out of which one with $Pc$ space group showed the largest polarization (around 90 $\mu C\ cm^{-2}$, Fig. S14).

Next, we took into account the mixing of Hf and Zr by optimizing those large-polarization structures with the four inequivalent ways to combine two Hf and two Zr atoms at the metal-atom locations. The resulting structures were very similar in formation energy, lattice parameters and oxygen positions (as seen in Fig. S14), indicating that the particular ordering at the cation sites is a secondary factor.

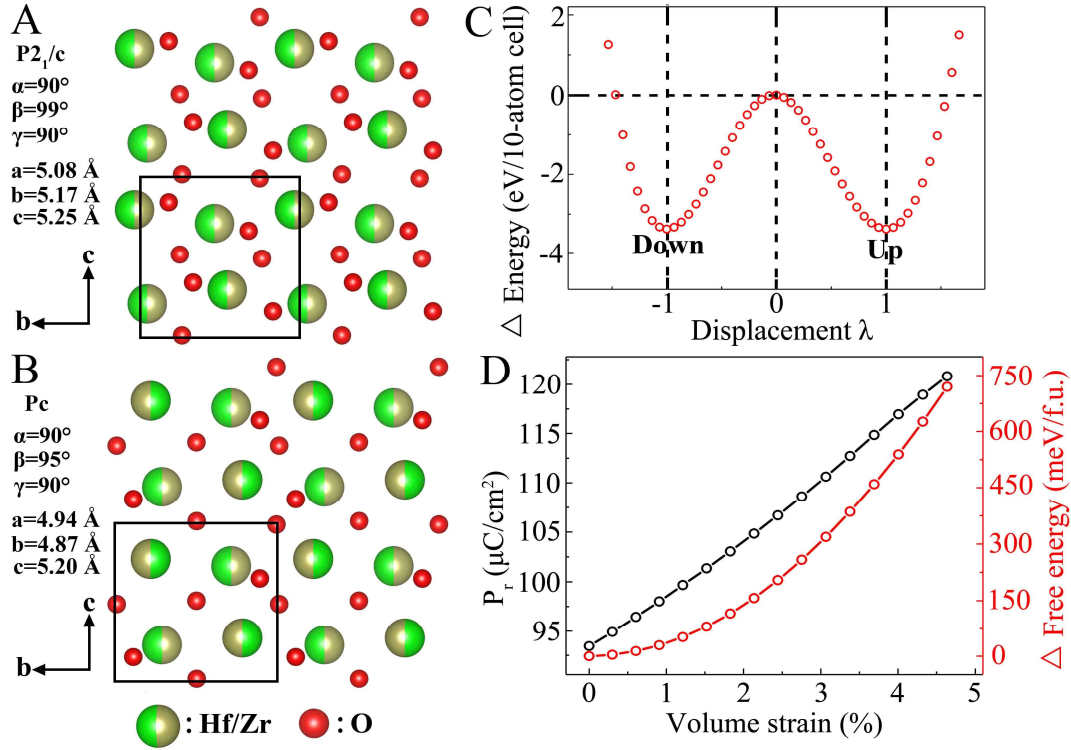

**Fig. 5. DFT calculations for proposed structure.** (**A**) Schematics of $P2_1/C$-phase structure model from standard CIF (*31*). (**B**) Schematics of constructed $Pc$-phase structure model from present DFT calculations of changing the cations into a 50%-50% species (**C**) Ferroelectric double-well energy landscape of strained HZO system with normalized displacements $\lambda$ of off-center oxygen atom. Calibrating the total free energy of paraelectric state of this HZO system to zero. (**D**) Calculated $P_r$ and free energy of the $Pc$ phase of this HZO system as a function of volume strain. Calibrating the total free energy of bulk state to zero.

We then selected one of those $Pc$-based structures and constrained its lattice vectors to take into account the strain conditions of the actual films. This resulted in $P_S$ growing up to around 120 μC cm$^{-2}$ (Figure 5D), very close to the value found experimentally at 10 K (our DFT calculations correspond to a temperature of 0 K). This strained structure is similar to the one observed by STEM images. The variation of the energy versus the normalized off-center oxygen atom displacements ($\lambda$) features the typical double-well landscape of displacive ferroelectrics (Figure 5C). The energy barrier between the paraelectric and ferroelectric structures is about 3.38 eV per unit cell when computed by interpolating between the found structure and its inverse—this is an upper bound on the height of the barrier, which would be lower if volume relaxation of the paraelectric structure was included, or if other possible low-energy paths were to exist.

In general, the presence of a giant depolarization field is expected in 10-nm-thickness films. Stabilization of the $P$c structure is critical for the generation of large ferroelectric polarization. On the one hand, since the bottom electrode (LSMO) could hardly provide perfect screening effect (even worse than $SrRuO_3$ (*32*)), the carrier screening mechanism generated by oxygen vacancies is crucial. A certain concentration of vacancies can generate a substantial number of carriers to screen the depolarization field in an effective way, which stabilizes the polar structure. On the other hand, the coexistence of multiple phases with laterally intertwined distribution of domains are likely to provide a highly strained state to the HZO thin films, which prompts the formation of $P$c phase and suppresses the growth of non-ferroelectric $P2_1/c$ phase during long-range annealing process.

**Conclusions**

We conclude that robust ferroelectric HZO thin films with ultralarge spontaneous polarization have been obtained by virtue of a long-range post-annealing process. A new ferroelectric monoclinic $P$c phase was identified, and confirmed by comprehensive experiments and theoretical calculations. These results provide direct evidence that the inherent polarization switching mainly results from the shifts of off-center oxygen atoms in the $P$c phase coexisting with the $P$ca$2_1$ phase. Our findings provide a paradigm shift in the field of ferroelectrics, as they demonstrate the potential for utilizing fluorite-structure materials with structure transition and defect engineering strategy, which may open an interesting route for designing high-performance, energy-efficient and environmentally friendly advanced devices in future electronics industries.

**Acknowledgments**

This research was supported by the National Key R&D Program of China (2020YFA0406202), the National Natural Science Foundation of China (22090042 and 21971009), Guangxi BaGui Scholars Special Funding, and the Fundamental Research Funds for the Central Universities, China (FRF-IDRY-GD21-03 and GJRC003). The authors gratefully acknowledge the cooperation of the beamline scientists at the Beijing Synchrotron Radiation Facility 1W1A and 4B9B beamline. M.C. and O.D. acknowledge funding from the Israel Science Foundation under Grant Number 3433/21 (ISF-NSFC).